\documentstyle[12pt]{article}
\begin{document}

\title{\Large\bf BV quantization of a vector-tensor gauge theory
with topological coupling}

\author{R. Amorim\thanks{\noindent Electronic mails: ift01001 @ ufrj
and amorim @ ifsu1.ufrj.br}~~ and
J. Barcelos-Neto\thanks{\noindent Electronic mails: ift03001
@ ufrj and barcelos @ vms1.nce.ufrj.br}\\
Instituto de F\'{\i}sica\\
Universidade Federal do Rio de Janeiro\\
RJ 21945-970 - Caixa Postal 68528 - Brasil}
\date{}

\maketitle
\abstract
We use the BV quantization method for a theory with coupled tensor
and vector gauge fields through a topological term. We consider in
details the reducibility of the tensorial sector as well as the
appearance of a mass term in the effective vectorial theory .

\vfill
\vspace{1cm}
\newpage

\section{Introduction}
\bigskip
The quantization method due to Batalin and Vilkoviski
(BV)~\cite{BV1,BV2} has been shown to be a powerful functional
procedure to deal with a wide variety of gauge theories. In the BV
scheme, covariance, reducibility, openness of gauge algebras and the
presence of gauge anomalies are taken into account in a very natural
way.

\medskip
In this work we consider the BV method to quantize the coupled
vector-tensor gauge theory present by Allen and
collaborators~\cite{Allen} and recently treated by
Lahiri~\cite{Lahiri} from the canonical point of view. It might be
opportune to mention that tensor gauge theories have attracted much
attention, formerly connected with the theory of strings, but also
with cosmic strings, vortices and black holes~\cite{tensor}. Another
use of tensorial gauge theories can be seen in connection with the
appearance of a mass term for vector gauge fields, in the context of
an effective theory~\cite{Lahiri}. As tensor gauge theories are
reducible, the BV scheme is a natural quantization procedure for
them.

\medskip
In the work of ref.~\cite{Lahiri}, it was not directly taken into
account the question of reducibility of constraints related to the
tensor gauge fields. This is an important point because if one use
that all the constraints are independent, the theory has zero tensor
degrees of freedom, what is not actually true~\cite{Kaul}. This point
is naturally considered here and the integration over the tensorial
sector can be done without infinities, leading to an effective
massive vectorial theory. We mention that in
references~\cite{Lahiri}, the presence of this mass term was
considered at classical level. Here, we investigate how it comes
quantically as a pole in the propagator of the gauge fixed effective
vector theory.

\medskip
Our work is organized as follows. In Sec.~2 we discuss the BV
quantization of a theory with topological coupling of vector and
tensor gauge fields. The obtainment of the effective mass for the
vector gauge field is achieved at Sec.~3. We leave Sec.~4 for some
concluding remarks.

\vskip1cm
\section {Vector coupled to tensor gauge fields}
\bigskip
The theory we are going to deal is described by the following
Lagrangian density~\cite{Allen,Lahiri}

\begin{equation}
{\cal L}=-\,\frac{1}{4}\,F_{\mu\nu}\,F^{\mu\nu}
-\,\frac{1}{12}\,H_{\mu\nu\rho}\,H^{\mu\nu\rho}\,
+{m\over4}\,\epsilon^{\mu\nu\rho\lambda}\,
B_{\mu\nu}F_{\rho\lambda}\,,
\label{2.1}
\end{equation}

\bigskip
\noindent where $F_{\mu\nu}$ and $H_{\mu\nu\rho}$ are totally
antisymmetric tensors written in terms of the potentials $A_{\mu}$
and $B_{\mu\nu}$ (also antisymmetric) through the curvature tensors

\begin{eqnarray}
F_{\mu\nu}&=&\partial_\mu A_{\nu}-\partial_\nu A_{\mu}\,,
\label{2.2}\\
H_{\mu\nu\rho}&=&\partial_\mu B_{\nu\rho}
+\partial_\rho B_{\mu\nu}+\partial_\nu B_{\rho\mu}\,.
\label{2.3}
\end{eqnarray}

\bigskip\noindent
In (\ref{2.1}),  $\epsilon^{\mu\nu\rho\lambda}$ is the totally
antisymmetric symbol and m is a mass parameter. It is easy to see, by
using the (coupled) Euler-Lagrange equations for $A^\mu$ and
$B^{\mu\nu}$, as well as the Jacobi identity, that $F_{\mu\nu}$
satisfy a massive Klein-Gordon equation, with mass parameter given by
the factor $m$ appearing in (\ref{2.1}).

\medskip
We observe that both  $F_{\mu\nu}$ and $H_{\mu\nu\rho}$ are invariant
under the gauge transformations

\begin{eqnarray}
\delta A^\mu&=&\partial^\mu\Lambda\,,
\label{2.4}\\
\delta B^{\mu\nu}&=&\partial^\mu\Lambda^\nu
-\partial^\nu\Lambda^\mu\,,
\label{2.5}
\end{eqnarray}

\bigskip\noindent
where $\Lambda$ and $\Lambda^\mu$ are (before fixing the gauge)
generic functions of spacetime. The reducible character of this
theory is seem from the fact that if we choose the gauge parameter
$\Lambda^\mu$ as the gradient of some scalar $\Omega$ we have that
$B^{\mu\nu}$ does not change under the gauge transformation
(\ref{2.5}). Actually, in this situation

\begin{eqnarray}
\delta B^{\mu\nu}&=&(\partial^\mu\partial^\nu
-\partial^\nu\partial^\mu)\,\Omega\,,
\nonumber\\
&=&0\,.
\label{2.6}
\end{eqnarray}

\bigskip
Expressions (\ref{2.4}) and (\ref{2.5}) can be rewritten as

\begin{eqnarray}
\delta A^\mu(x)&=&\int d^4y\,R^\mu(x,y)\,\Lambda(y)\,,
\label{2.7}\\
\delta B^{\mu\nu}(x)&=&\int d^4y\,
R^{\mu\nu}_\rho(x,y)\,
\Lambda^\rho(y)\,,
\label{2.8}
\end{eqnarray}

\bigskip\noindent
where

\begin{eqnarray}
R^\mu(x,y)&=&\partial^\mu\,\delta(x-y)\,,
\label{2.9}\\
R^{\mu\nu}_\rho(x,y)&=&\partial\,^{[\mu}\delta(x-y)\,
\delta^{\nu]}_\rho
\label{2.10}
\end{eqnarray}

\bigskip\noindent
are the generators of the gauge symmetries, and

\begin{equation}
Z^\mu(x,y)=\partial^\mu\,\delta(x-y)
\label{2.11}
\end{equation}

\noindent\bigskip
is the reducibility operator~\cite{BV1,BV2} corresponding
to~(\ref{2.6}) since

\begin{equation}
\int d^4y\,Z^\rho(x,y)\,R^{\mu\nu}_\rho(y,z)=0\,.
\label{2.11a}
\end{equation}

\bigskip\noindent
In expression~(\ref{2.10}), as well as in some forthcoming ones, the
antisymmetrization is performed in a normalized way, i.e.
$K^{[\mu\nu]}=(K^{\mu\nu}-K^{\nu\mu})/2$ for some $K^{\mu\nu}$
tensor.

\medskip
We are now ready to write down the solution of the classical master
equation. From expressions~(\ref{2.9}~-~\ref{2.11}) we notice that the
algebra of the gauge generators is closed and Abelian. To fix the
notation, it is convenient to distinguish between both sectors
(vectorial and tensorial) of the theory. Let us so generically denote
by $c$ and $d$ the ghosts associated respectively to the vectorial
and to the tensorial sectors.  In this way, the solution we are
looking for reads~\cite{BV1,BV2}

\begin{eqnarray}
S&=&S_0+\int d^4xd^4y\,\Bigl[i\,A^*_\mu(x)\,
\partial^\mu_x\,\delta(x-y)\,c(y)
\nonumber\\
&&\phantom{S_0}+2\,i\,B^*_{\mu\nu}\,\partial_x^{[\mu}\,
\delta(x-y)\,\delta^{\nu]}_\rho\,d^\rho(y)
+d^*_\rho(x)\,\partial^\rho_x\,\delta(x-y)\,d(y)\Bigr]\,,
\nonumber\\
&=&S_0+\int d^4x\,\Bigl[i\,A^*_\mu\,\partial^\mu c
+i\,B^*_{\mu\nu}\,\partial^{\mu}d^{\nu}
+d^*_\mu\,\partial^\mu d\Bigr]\,,
\label{2.12}
\end{eqnarray}

\bigskip\noindent
where $S_0$ is the action corresponding to the Lagrangian~(\ref{2.1})
and $A^\ast_\mu$, $B^\ast_{\mu\nu}$, $c^\ast_\mu$, $d^\ast_\mu$ are
antifields of the BV formalism.

\medskip
To fix the gauge freedom of the vectorial sector, we introduce the
gauge-fixing fermionic functional

\begin{equation}
\psi=-\int d^4x\,\,\bar c\,\Bigl(\partial^\mu A_\mu
-\frac{\alpha}{2}\,b\Bigr)+\cdots\,.
\label{2.13}
\end{equation}

\bigskip\noindent
and extend the classical action to

\begin{equation}
\bar S=S+\int d^4x\,\bar c^*\,b+\cdots
\label{2.14}
\end{equation}

\bigskip\noindent
Here, dots are representing the corresponding quantities for the
tensorial sector which we are going to discuss soon. After
substituting the antifields $A_\mu^\ast$ and $\bar c^\ast$ by the
derivatives of $\psi$ with respect to $A_\mu$ and $\bar c$ we arrive,
for the vectorial sector, at the usual Faddeev-Popov expression for
the vacuum functional with covariant Gaussian gauge-fixing.

\medskip
The fixation of the tensorial counterpart is not so simple. First of
all we need to have a covariant gauge fixing with the same
reducibility character of the gauge freedom of $H_{\mu\nu\rho}$. It
may be chosen as

\begin{equation}
\Xi^\nu=\partial_\mu\,B^{\mu\nu}\,.
\label{2.15}
\end{equation}

\bigskip\noindent
We notice that $\Xi^\mu$ is divergenceless. To implement~(\ref{2.15})
we first observe that in~(\ref{2.13}) and (\ref{2.14}) we have used
the ghosts $c$, $c^*$, $\bar c$ and $\bar c^*$ to fix the vectorial
gauge invariance of~(\ref{2.1}). As we have already used the unbarred
quantities $d^\mu$, $d_\mu^*$, $d$ and $d^*$ in the tensorial sector
of~(\ref{2.12}), we expect to use the additional pairs $\bar d_\mu$,
$\bar d^{*\mu}$, $\bar d$ and $\bar d^*$ in a similar fashion. This
gives the tensorial sector of the BV action and the corresponding
fixing fermion functional

\begin{eqnarray}
\bar S&=&\cdots\,+\int d^4x\,\bar d^*_\mu\,e^\mu\,,
\label{2.16}\\
\psi&=&\cdots\,-\int d^4x\,\Bigl[\bar d_\mu\,
\Bigl(\partial_\nu B^{\nu\mu}
-{\beta\over2}\,e^\mu\Bigr)
+\bar d\,\partial_\mu\,d^\mu\Bigr]\,,
\label{2.17}
\end{eqnarray}

\bigskip\noindent
where $e^\mu$ is an auxiliary field that plays a similar role as $b$.
As we can observe, the reducibility of the original theory has not
been completely fixed yet. Even though the ghost gauge invariance

\begin{equation}
d^\mu\longrightarrow d^\mu +\partial^\mu\,\zeta
\label{2.18}
\end{equation}

\bigskip\noindent
has been fixed in~(\ref{2.17}), there remains to consider the further
invariance

\begin{equation}
\bar d^\mu\longrightarrow\bar d^\mu +\partial^\mu\,\bar\zeta
\label{2.19}
\end{equation}

\bigskip\noindent
of the complete action. This has already been discussed by Henneaux
and Teitelboim in~ref.\cite{BV2}.  The solution comes by introducing
further pairs ($\eta$,~$\eta^*$) and ($f$,~$f^*$) in the theory. This
is achieved by adding the term $-\int d^4x\,\eta\,\partial^\mu\,\bar
d_\mu$ to the fixing fermion $\psi$ and $i\int d^4x\,\eta^*f$ to the
action. In these expressions, $f$ can be considered as the ghost
corresponding to symmetry~(\ref{2.19}) and $\eta$ can be seen as an
auxiliary field, on the same footing as $b$ and $b^\mu$.

\medskip
Putting the considerations above all together, we can write the
complete BV action as

\begin{eqnarray}
\bar S=S_0&\!\!+\!\!&\int d^4x\,
\Bigl(i\,A^*_\mu\,\partial^\mu c+\bar c^*\,b
+i\,B^*_{\mu\nu}\,\partial^\mu d^\nu
\nonumber\\
&\!\!+\!\!&d^*_\mu\,\partial^\mu d+\bar d^*_\mu\,e^\mu
+i\,\bar d^*\,\bar f+i\,\eta^*f\Bigr)
\label{2.20}
\end{eqnarray}

\bigskip\noindent
and the complete gauge fixing fermion functional as

\begin{eqnarray}
\psi=-\int d^4x\,\Bigl[\bar c\,
\Bigl(\partial_\mu\,A^\mu -{\alpha\over 2}\,b\Bigl)
&+&\bar d_\mu\,\Bigl(\partial_\nu\,B^{\nu\mu}
-{\beta\over2}\,e^\mu\Bigr)
\nonumber\\
&+&\bar d\,\partial_\mu\,d^\mu
+\eta\,\partial^\mu\,\bar d_\mu\Bigr]\,.
\label{2.21}
\end{eqnarray}

\bigskip
At this stage, it might be elucidative to present a table of parity
and ghost numbers for fields and antifields of the theory.

\begin{eqnarray}
\epsilon\,[A^\mu,B^{\mu\nu},b,d,\bar d,f^*,\bar f^*,e^\mu,c^*,\bar c^*,d^*_\mu,
\bar d^{*\mu},\eta]&=&0\,,
\nonumber\\
\epsilon\,[A^*_\mu,B^*_{\mu\nu},b^*,d^*,\bar d^*,f,\bar f,e^*_\mu,
c,\bar c,d^\mu,\bar d_{\mu},\eta^*]&=&1\,,
\label{2.22}\\
\nonumber\\
gh(d^*)&=&-3\,,
\nonumber\\
gh(c^*,d^*_\mu,\bar d,f^*)&=&-2\,,
\nonumber\\
gh(A^*_\mu,B^*_{\mu\nu},\bar c,b^*,\bar d_\mu,e^*_\mu,\eta^*,\bar f)&=&-1\,,
\nonumber\\
gh(A^\mu,B^{\mu\nu},\bar c^*,b,\bar d^{*\mu},e^\mu,\eta,\bar f^*)&=&0\,,
\nonumber\\
gh(c,d^\mu,\bar d^*,f)&=&1\,,
\nonumber\\
gh(d)&=&2\,.
\label{2.23}
\end{eqnarray}

\bigskip\noindent
Making use of the tables above, we observe that action~(\ref{2.20})
actually has even Grassmamnian parity and ghost number zero, as
expected.

\medskip
With the gauge-fixing functions, antifields are obtained as usual

\begin{eqnarray}
A_\mu^*&=&{\delta \psi\over{\delta A^\mu }}
\,\,=\,\,\partial_\mu\bar c\,,
\nonumber\\
B_{\mu\nu}^*&=&{\delta\psi\over{\delta B^{\mu\nu}}}
\,\,=\,\,\partial\,_{[\mu}\,\bar d_{\nu]}\,,
\nonumber\\
\bar c^*&=&{\delta\psi\over{\delta\bar c}}
\,\,=\,\,-\,\partial_\mu A^\mu
+{\alpha\over2}\,b\,,
\nonumber\\
d_\mu^*&=&{\delta\psi\over{\delta d^\mu}}
\,\,=\,\,\partial_\mu\,\bar d\,,
\nonumber\\
\bar d^{*\mu}&=&{\delta\psi\over{\delta\bar d_\mu}}
\,\,=\,\,-\,\partial_\nu B^{\nu\mu}+{\beta\over2}\,e^\mu+\partial_\mu\eta\,,
\nonumber\\
\bar d^*&=&{\delta\psi\over{\delta\bar d}}
\,\,=\,\,-\,\partial_\mu\,d^\mu\,,
\nonumber\\
\eta^*&=&{\delta \psi\over{\delta\eta}}
\,\,=\,\,-\,\partial^\mu\,\bar d_\mu  \,.
\nonumber\\
b^*&=&{\delta \psi\over{\delta b}}
\,\,=\,\,{\alpha\over2}\,\bar c \,.
\nonumber\\
e_\mu^*&=&{\delta \psi\over{\delta e^\mu}}
\,\,=\,\,{\beta\over2}\,\bar d_\mu  \,.
\label{2.24}
\end{eqnarray}

\bigskip\noindent
Now, the combination of~(\ref{2.23}) and~(\ref{2.24}) leads to

\begin{eqnarray}
\bar S\,\,=\,\,S_0&\!+\!&\int d^4x\,
\Bigl[i\,\partial_\mu\bar c\,\partial^\mu c
+\Bigl(\partial_\mu A^\mu-{\alpha\over2}\,b\Bigr)\,b
+i\,\partial\,_{[\mu}\bar d_{\nu]}\,
\partial^{\mu} d^{\nu}
\nonumber\\
&\!+\!&\partial_\mu\bar d\partial^\mu d
+\Bigl(\partial_\nu B^{\nu\mu}-{\beta\over2}\,e^\mu
-\partial^\mu\eta\Bigr)\,e_\mu
-i\,\partial_\mu d^\mu\bar f + i\,f\partial^\mu\bar d_\mu\Bigr]\,.
\label{2.25}
\end{eqnarray}

\bigskip
Once the process of construction of the gauge fixed BV action is done
and as we know that the theory is free of anomalies~\cite{BV2}, the
next step for the BV quantization is to write the vacuum functional
(or its external current dependent generalizations)

\begin{equation}
Z=\int\,[d\mu]\,\exp\{i\bar S\}\,,
\label{2.26}
\end{equation}

\noindent where $\bar S$ is given by~(\ref{2.25}) and $[d\mu]$
represents the Liouville measure for all the fields appearing there.
If we introduce external currents, we are able to generate all the
Green's functions of the theory in a perturbative scheme, by the
usual derivative expansion in the external sources. An interesting
point we are going to discuss in the next section is how the
functional integrations over the antisymmetric tensor field can be
performed in a closed way, leading to an effective massive theory for
the vectorial field.

\vskip 1cm
\section{Massive vectorial effective theory}
\bigskip
As we have mentioned in the last section, the classical equations of
motion for the fields $A^\mu$ and $B^{\mu\nu}$ can be manipulated in
order to show that $F^{\mu\nu}$ satisfies a massive Klein-Gordon
equation.  The same kind of feature could be obtained for
$H^{\mu\nu\rho}$, after eliminating  $F^{\mu\nu}$. We can make two
comments: the first one is that these results are obviously
classical. The second one is that are not the fields themselves that
satisfy massive Klein-Gordon equations, but their curvature tensors.
To see how corresponding features appear at quantum level, we are
going to obtain an effective theory, first by functionally
integrating over the auxiliary fields $b$, $e^\mu$  and $\eta$ to introduce
Gaussian fixing terms in the action . After that, we get for the
$B^{\mu\nu}$ depending part of the gauge-fixed effective action

\begin{eqnarray}
\bar S_B&=&\int d^4x\,\Bigl[-\,\frac{1}{12}\,
H_{\mu\nu\rho}\,H^{\mu\nu\rho}\,
-{1\over{2\beta}}\,\bigl(\partial_\nu B^{\nu\mu}\bigr)^2
+{m\over4}\,\epsilon^{\mu\nu\rho\lambda}\,
B_{\mu\nu}F_{\rho\lambda}\Bigr]\,,
\nonumber\\
&=&\int d^4x\,\Bigl[{1\over4}\,B_{\nu\lambda}\,
O^{\nu\lambda}_{\alpha\beta}\,B^{\alpha\beta}
+{1\over2}\,D^{\nu\lambda}\,B_{\nu\lambda}\Bigr]\,.
\label{3.1}
\end{eqnarray}

\bigskip\noindent In order to simplify the notation we have defined
the operator

\begin{equation}
O^{\mu\nu}_{\alpha\beta}={1\over2}\,
\delta^{\mu\nu\lambda}_{\alpha\beta\gamma}\,
\partial_\lambda\,\partial^\gamma
+{2\over\beta}\,\delta\,^{[\mu}_{[\alpha}
\partial^{\nu]}\partial_{\beta]}
\label{3.2}
\end{equation}

\noindent and the dual tensor

\begin{equation}
D^{\mu\nu}={m\over2}\,\epsilon^{\mu\nu\rho\lambda}\,
F_{\rho\lambda}\,.
\label{3.3}
\end{equation}

\bigskip\noindent
In expression~(\ref{3.2}), $\delta^{\mu\nu\lambda}_{\alpha\beta\gamma}=6\,
\delta\,^\mu_{[\alpha}\delta^\nu_\beta\delta^\lambda_{\gamma]}$ values 1
if $\mu,\nu,\lambda$ are an even permutation of $\alpha,\beta,\gamma$.

\medskip
To perform the integral in the $B$'s we need the inverse of
$O^{\mu\nu}_{\alpha\beta}$ given by~(\ref{3.2}), in such a way that

\begin{equation}
O^{\mu\nu}_{\alpha\beta}\,(O^{-1})_{\mu\nu}^{\rho\sigma}
=\delta_{[\alpha}^\rho\,\delta_{\beta]}^\sigma\,.
\label{3.4}
\end{equation}

\bigskip\noindent
It is just a matter of algebraic calculation to show that

\begin{equation}
(O^{-1})_{\mu\nu}^{\rho\sigma}={1\over2\Box}\,
\Bigl[\delta_{\mu\nu}^{\rho\sigma}
-{4\over\Box}\,\Bigl(1-{\beta\over2}\Bigr)\,
\delta_{[\mu}^{[\rho}\partial_{\nu]}\partial^{\sigma]}\Bigr]\,.
\label{3.5}
\end{equation}

\bigskip\noindent
With the aid of~(\ref{3.5}), we can see that after functionally
integrating over $B^{\mu\nu}$, the part of the effective action which
had depended of the tensorial field acquires the form

\begin{eqnarray}
\bar S_B&=&-\,{1\over4}\int d^4x\,D^{\mu\nu}\,
(O^{-1})_{\mu\nu}^{\rho\sigma}\,D_{\rho\sigma}\,,
\nonumber\\
&=&-\,{1\over4}\int d^4x\,D_{\mu\nu}\,
\Bigl[{1\over\Box}\,\delta^\nu_\sigma
-{2\over{\Box^2}}\,\Bigl((1-{\beta\over2}\Bigr)\,
\partial_\sigma\partial^\nu\Bigr]\,D^{\mu\sigma}\,.
\label{3.6}
\end{eqnarray}

\bigskip\noindent
By using the explicit form of $D_{\mu\nu}$ given by~(\ref{3.3}) and
the Jacobi identity satisfied by $F_{\mu\nu}$, we observe that the
term proportional to $(1-{\beta\over2})$ in~(\ref{3.6}) vanishes
identically. The final result reads

\begin{eqnarray}
\bar S_B&=&-\,{m^2\over4}\int d^4x\,F_{\mu\nu}\,
{1\over\Box}\,F^{\mu\nu}\,,
\nonumber\\
&=&-\,{m^2\over2}\int d^4x\,\Bigl(A_\mu\,A^\mu
+{\partial_\mu A^\mu\over\Box}\Bigr)\,
\partial_\nu A^\nu\,.
\label{3.7}
\end{eqnarray}

\bigskip\noindent
At last we have that the complete effective action can be written as

\begin{equation}
\bar S=\int d^4x\,\Bigl[{1\over2}\,A_\mu\,
\bigl(\Box - m^2\bigr)\,A^\mu
-\,{1\over2}\,\partial_\mu\,A^\mu\,
\Bigl((1-{1\over\alpha} -{m^2\over\Box}\Bigr)\,
\partial_\nu\,A^\nu\Bigr] + S_{ghost}\,,
\label{3.8}
\end{equation}

\bigskip\noindent
where $S_{ghost}$ represents the ghost dependent part of the
effective action which appears in the vacuum functional~(\ref{2.26}).

\medskip
The operator which acts on (the quadratical part of) $A^\mu$
in~(\ref{3.7}), written in momentum space, reads

\begin{equation}
G_{\mu\nu}=-\,\Bigl[\bigl(k^2+m^2)\,\eta_{\mu\nu}
-\,\Bigl(1-{1\over\alpha}+{m^2\over k^2}\Bigr)\,
k_\mu k_\nu\Bigr]\,.
\label{3.9}
\end{equation}

\bigskip\noindent
Its inverse gives the propagator  for the vectorial field in momentum
space. It is just given by

\begin{equation}
K_{\mu\nu}=-\frac{1}{k^2+m^2}\,
\Bigl[\eta_{\mu\nu}+\Bigl(\frac{\alpha-1}{k^2}+
\frac{m^2}{k^4}\Bigr)\,k_\mu k_\nu\Bigr]\,,
\label{3.10}
\end{equation}

\bigskip\noindent
which presents a pole in $k_0=\vec k^2+m^2$, showing that at quantum
level the elimination of the tensorial fields is also traduced by the
introduction of a mass term for the vectorial field.  Observe however
that here it becomes clear that is the vectorial field itself that
effectively acquires a mass.

\vskip1cm
\section{Conclusion}
\bigskip
With the aid of the Batalin and Vilkoviski procedure,
we have constructed the vacuum functional for a field theoretical
model consisting of vector and tensor gauge fields interacting through
a topological term. The reducibility of the tensorial sector was taken
properly into account. After a covariant gauge-fixing was implemented,
we were able to integrate over the tensorial fields, obtaining
a massive vectorial effective theory  as output. The propagator
for the vectorial bosons have been calculating, showing the expected
pole at $\vec p\,^2+m^2$.

As a final comment, we observe that the introduction of non-Abelian
gauge symmetries is possible for a corresponding extended
model~\cite{Bar}.  Its BV formulation, as well as some
phenomenological related consequences for this non-Abelian version of
the present model are under study and will be presented
elsewhere~\cite{BR}.

\vskip 1cm
\noindent {\bf Acknowledgment:}
This work is supported in part by
Conselho Nacional de Desenvolvimento Cient\'{\i}fico e Tecnol\'ogico
- CNPq (Brazilian Research Agency).

\end{document}